\documentclass[a4paper,10pt,twocolumn]{revtex4}
\usepackage{fontenc}
\usepackage[spanish]{babel}
\usepackage{textcomp}
\usepackage{amsfonts}
\usepackage{amsmath}
\usepackage{amssymb}
\usepackage{amsthm}
\usepackage{fontenc}
\usepackage[pdftex]{graphicx}
\usepackage{subfigure}
\usepackage{upgreek}
\usepackage[utf8]{inputenc}
\usepackage[dvips,pdftex]{hyperref}

\begin{document}

\title{Nambu brackets for the electromagnetic field}
\author{Roberto Salazar }
\affiliation{Center for Optics and Photonics, University of Concepción, Casilla 4016, Concepción Chile.
  }
\author{Michael V. Kurgansky}
\affiliation{A.M. Obukhov Institute of Atmospheric Physics, Pyzhevsky 3 119017 Moscow, Russia.}
\date{November 24, 2010}

\begin{abstract}
A Nambu formulation for the electromagnetic field in the case of stationary charge density and vanishing charge current density is proposed.

\end{abstract}

\maketitle
\section{Introduction}
In Hamiltonian mechanics the governing equations are obtained from the Poisson brackets formalism. Poisson brackets formalism gives a prescription to find time derivative of any function $F$ using the Hamiltonian function $H$ and a bilinear antisymetric function constructed using the derivatives with respect to the pairs of phase space coordinates $(p_k,q_k)$, $\quad k = 1,...,N$ known as the Poisson bracket (PB) $[.,.]_{p_k,q_k}$ i.e.:
\begin{equation}
 \frac{dF}{dt} = \sum_k[F,H]_{p_k,q_k}
\end{equation}
Nambu mechanics \cite{Nambu} is a novel generalization of Hamiltonian mechanics where the prescription to  find the time derivative of any function $F$ uses two conserved quantities $I_1$, $I_2$ and a trilinear completely antisymmetric combination of the derivatives with respect to the extended phase space triplets of coordinates  $(x_k,y_k,z_k)$, $\quad k = 1,...,N$ known as the Nambu bracket (NB):

\begin{equation}
 \frac{dF}{dt} = \sum_k[F,I_1,I_2]_{x_k,y_k,z_k}
\end{equation}

Nambu brackets conserve two or more quantities by construction, so for a physical system with a finite number of conserved quantities a Nambu description has less degeneracies than a Poisson description. In this sense we say that Nambu  brackets are less degenerate than the Poisson bracket. Névir and Blender \cite{NevBle}  established a Nambu field approach to ideal fluid mechanics, and the above-mentioned less degeneracy property allowed  Salmon \cite{Salmon1,Salmon2} to make a first application of this theory. He established that the discretization of the Nambu brackets leads to numerical schemes that conserve energy and vorticity related quantities. He showed that the Arakawa Jacobian can be derived systematically from the antisymetry of the Nambu representation. \\

In a recent paper the authors \cite{RobKur} proposed a classification of Nambu brackets where  Nambu brackets of first kind (NB I)  are those   NBs  or sums of NBs which  involve the same conserved quantities \cite{Nambu}. The sums of NBs that involve different conserved or, alternatively, constitutive  quantities are called the Nambu brackets of second kind (NB II) \cite{Nambu}. Like in \cite{Nevir} we call the constitutive quantities those ones which form the basis  of Nambu formalism but are not necessarily the constants of motion. \\

Our main result is an extension of Nambu formalism onto the electromagnetic field in the case of constant electric charge density and vanishing electric current density by using a NB I that involves two conserved quantities neither of them being the energy. 

\section{Electromagnetism}

Electromagnetic theory is one of the most fundamental physical theories  with applications in almost all areas of physics, ranging from atomic interactions to astrophysical phenomena. The governing equations of electromagnetism are the famous Maxwell equations, which  in Lorentz-Heaviside units read:
\begin{align}
 \nabla\cdot \mathbf{E} &= \rho \\
 \nabla\cdot\ \mathbf{B}  &= 0 \\
 \frac{\partial \mathbf{B}}{\partial t} &= -\nabla\times\mathbf{E} \\
  \frac{\partial \mathbf{E}}{\partial t} &= \nabla\times\mathbf{B} -\mathbf{J}
\end{align}
and the continuity equation:
\begin{equation}
 \frac{\partial \rho}{\partial t} = -\nabla\cdot\mathbf{J}
\end{equation}

Here $\mathbf{E}$ is the electric field, $\mathbf{B}$ is the magnetic flux density field, $\rho$ the electric charge density and $\mathbf{J}$ is the electric current density. This is a set of linear  partial differential equations, but applying a Fourier transform we can obtain a system of formally  non-linear ordinary differential equations, where the wavenumber vector $\mathbf{k}$ becomes a variable:
\begin{align}
 \mathbf{k}\cdot \mathbf{\tilde{E}} &= \tilde{\rho} \\
 \mathbf{k}\cdot\ \mathbf{\tilde{B}}  &= 0 \\
 \frac{\partial \mathbf{\tilde{B}}}{\partial t} &= -\imath\mathbf{k}\times\mathbf{\tilde{E}} \\
  \frac{\partial \mathbf{\tilde{E}}}{\partial t} &= \imath\mathbf{k}\times\mathbf{\tilde{B}} -\mathbf{\tilde{J}}\\
 \frac{\partial \tilde{\rho}}{\partial t} &= -\imath\mathbf{k}\cdot\mathbf{\tilde{J}}
\end{align}
The $\sim$ means that the variable is now a function of $\mathbf{k}$ in Fourier space. We consider the case when $\tilde{\rho}$ does not depend on time and $\mathbf{\tilde{J}} =0$. From this we see that the electric charge conservation equation is satisfied identically. We consider equations (8) and (9) as constraints, so we proposed for any functional $\mathcal{F} = \mathcal{F}(\mathbf{\tilde{E},\tilde{B}})$ the following NB I:
\begin{equation}
 \frac{d\mathcal{F}}{d t} = [\mathcal{F},\mathcal{I}_1,\mathcal{I}_2]_{\mathbf{\tilde{E}},\mathbf{\tilde{E}},\mathbf{\tilde{E}} } +  [\mathcal{F},\mathcal{I}_1,\mathcal{I}_2]_{\mathbf{\tilde{B}},\mathbf{\tilde{B}},\mathbf{\tilde{B}} }
\end{equation}

 $\mathcal{I}_j$ , $j=1,2$ are the invariants:

\begin{align*}
 \mathcal{I}_1 &=\int d^{3}k \{ \mathbf{k}\cdot\ \mathbf{\tilde{E}} - \mathbf{k}\cdot \mathbf{\tilde{B}}\} \\
 \mathcal{I}_2 &= \int d^{3}k \{\mathbf{\tilde{E}}\cdot\mathbf{\tilde{B}} \}
\end{align*}
and the Nambu brackets are of the form: 

\begin{align}
 [\mathcal{F},\mathcal{I}_1,\mathcal{I}_2]_{\mathbf{\tilde{B}},\mathbf{\tilde{B}},\mathbf{\tilde{B}} } &= \imath\int d^{3}k \frac{\delta \mathcal{F}}{\delta\mathbf{\tilde{B}} } \cdot\left(\frac{\delta \mathcal{I}_1}{\delta\mathbf{\tilde{B}} }\times\frac{\delta \mathcal{I}_2}{\delta\mathbf{\tilde{B}} } \right)\\
 [\mathcal{F},\mathcal{I}_1,\mathcal{I}_2]_{\mathbf{\tilde{E}},\mathbf{\tilde{E}},\mathbf{\tilde{E}} } &= \imath\int d^{3}k \frac{\delta \mathcal{F}}{\delta\mathbf{\tilde{E}} } \cdot\left(\frac{\delta \mathcal{I}_1}{\delta\mathbf{\tilde{E}} }\times\frac{\delta \mathcal{I}_2}{\delta\mathbf{\tilde{E}} } \right)
\end{align}

This construction conserves directly the global electric charge $\mathcal{I}_1 $ (see equations (8),(9)) and one of the classical electromagnetic invariants $\mathcal{I}_2$, but not the energy $\mathcal{H} = \frac{1}{2}\int d^{3}k \{  |\mathbf{\tilde{E}}|^{2} + |\mathbf{\tilde{B}}|^{2}\}$ or the other classical electromagnetic invariant: $\int d^{3}k \{  |\mathbf{\tilde{E}}|^{2} -  |\mathbf{\tilde{B}}|^{2}\}$. Nevertheless the latter quantities are still conserved due to equations (8)-(12) and because of that  they are degeneracies of the Nambu bracket. Like in \cite{Nevir} we say that this kind of quantities are Supercasimirs.\\

  Using this construction we are extending Nambu's original proposal to describe mechanics within a formalism where energy doesn't have a superior hierarchy over other conserved quantities (casimirs). We have shown that this is possible for electromagnetism, and in a way that energy takes a merely supercasimir status.\\
It is an interesting feature of this construction that we can construct the Nambu bracket when we apply Fourier transform to
the original Maxwell equations and  their conservation laws i.e., by changing from linear partial differential equations to non-linear ordinary differential equations.\\
It is also interesting to notice that using the generalized Jacobi identity for Nambu mechanics \cite{Cari}, we can find a novel non-trivial invariant $\mathcal{G}$ for this system using as basis the invariants $\mathcal{I}_1$, $\mathcal{I}_2$ and $\mathcal{H}$: 
\begin{equation}
 \mathcal{G} =\int d^{3}k \{\mathbf{k} \cdot(\mathbf{\tilde{E}}\times\mathbf{\tilde{B}})\}
\end{equation}

\section{Concluding Remarks}

We have constructed a Nambu formalism for the Maxwell equations in the case of stationary electric charge density and vanishing electric current. This formalism was found by transforming Maxwell's linear partial differential equations to Fourier space where they are non-linear ordinary differential equations. Also this formalism conserves energy as a supercasimir and allows us to find a non-trivial invariant by use of the generalized Jacobi identity. We want to notice that by doing this in electromagnetism we have also done it for gravitomagnetism since the latter equations are  isomorphous to the Maxwell equations. So, the obtained result has also implication for the general relativity in the limit case of weak gravitational fields \cite{Clark}.

  A remarkably fact is the existence of a deduction of  Maxwell equations in empty space from pure mechanical considerations. This deduction is due to MacCullagh in 1839 (see \cite{Sommerfeld}) by means of a hypothetical quasi-elastic body responsive to rotations relative to absolute space. Since Nambu formalism is relative to absolute space and we made the construction by means of a Fourier transformation we think that both deductions are close, in the sense that both allow us to give a mechanical interpretation of Maxwell equations in empty space. Of course, this is only a mathematical curiosity.


\begin{thebibliography}{References}

\bibitem{Nambu} 
Nambu, Y. 1973: Generalized Hamiltonian dynamics. \textit{Phys. Rev. D, \textbf{7}, 2405-2412}.
\bibitem{NevBle} Névir P. and Blender 1993: A Nambu representation of incompressible hydrodynamics using helicity and enstrophy \textit{J. Phys A, \textbf{26},L1189-L1193 }.
\bibitem{Salmon1} Salmon R. 2005: A general method for conserving quantities related to potencial vorticity in numerical models, \textit{Institute of Physics Publishing, Nonlinearity,\textbf{18} R1-R16}. 
\bibitem{Salmon2} Salmon R. 2007: A general method for conserving energy and potential enstrophy in shallow-water models \textit{Journal of the Atmospheric Sciences, \textbf{64}, 515-530}.
\bibitem{Nevir} Névir P. and Sommer M. 2009: Energy-Vorticity theory of ideal fluid mechanics \textit{Journal of the Atmospheric Sciences, \textbf{66}, 2073-2084}.
\bibitem{Cari} Cari\~nena J.,Guha P. and Ra\~nada M. 2008: Hamiltonian and quasi-Hamiltonian systems, Nambu-Poisson structures and symmetries \textit{J. Phys. A: Math. Theor. \textbf{41}, 335209}
\bibitem{RobKur} Salazar R. and Kurgansky M. 2010: Nambu brackets in fluid mechanics and magnetohydrodynamics \textit{ J. Phys. A: Math. Theor. \textbf{43}, 305501}
\bibitem{Clark}  Clark S. and Tucker R. 2000: Gauge Symmetry and gravito-electromagnetism \textit{J. Phys. Class. Quantum Grav. \textbf{17}, 4125}
\bibitem{Sommerfeld} Sommerfeld A."Mechanics of Deformable Bodies", Lectures on Theoretical Physics, Volume II, Academic press, 1964.
\end{thebibliography}
\end{document}